\documentclass[runningheads]{llncs}
\usepackage{graphicx}
\usepackage{wrapfig}
\usepackage{float}
\usepackage{amsmath}
\usepackage{amssymb}
\usepackage{mathtools}
\usepackage[english]{babel}
\usepackage{cite}
\usepackage{textcmds}
\usepackage[utf8]{inputenc}
\usepackage{lineno}
\usepackage{dirtytalk}
\usepackage{xspace}
\usepackage{wrapfig}
\usepackage{enumitem}

\newcommand{\Heraklit}{\normalfont \textsc{Heraklit}\xspace}

\newcommand\sbullet[1][.75]{\mathbin{\vcenter{\hbox{\scalebox{#1}{$\bullet$}}}}}

\DeclareMathOperator{\compose}{\sbullet}

\newcommand{\defeq}{\mathrel{=_{\text{def}}}}

\begin{document}
\title{Trust by design -- in praise of modularization:\\a case study}
%
%

\author{Peter Fettke\inst{1,2}\orcidID{0000-0002-0624-4431} \and
Wolfgang Reisig\inst{3}\orcidID{0000-0002-7026-2810}}
\authorrunning{P. Fettke, W. Reisig}
%

\institute{German Research Center for Artificial Intelligence (DFKI), Saarbr\"ucken, Germany \\
\email{peter.fettke@dfki.de}\\ \and
Saarland University, Saarbr\"ucken, Germany \\ \and
Humboldt-Universität zu Berlin, Berlin, Germany \\ 
\email{reisig@informatik.hu-berlin.de}}

\maketitle 
\begin{abstract}
Ensuring that collective adaptive systems remain safe, reliable, and trustworthy requires measures that transcend so far established formal methods, and in particular established verification techniques. In this contribution, we suggest three such measures: (1) conceptual means: runs with locally confined cause and effect of events, (2) temporal logic like verification techniques that respect and exploit such runs, (3) composing system properties from properties of components. This contribution presents a case study which particularly focuses on the benefits of modularization for achieving trust by design. Further work will develop a full-fledged theory for the presented ideas.

\end{abstract}

\section{Introduction}\label{sec:1}

In collective adaptive systems, the interplay between software components, AI-based techniques, and human actors can produce emergent behaviors that are challenging to predict and analyze. Ensuring that these systems stay safe, reliable, and trustworthy requires methods that go beyond currently established formal methods and verification techniques.

In this paper, we highlight \emph{modularity} as a key to creating trustworthy systems and system models. Not only should systems be built from modules, but each single system run and each correctness argument should also follow this concept. Conceptually, all models of systems, runs, and correctness arguments share the same kind of modular structure of modules and their composition. They intertwine, yielding easy-to-comprehend models. 

Here, we do not present a fully developed theory of modular system construction, but illustrate the decisive concepts and their harmonic interplay through a case study. Section~\ref{sec:2} presents the case study informally: the refund department of a company. Section~\ref{sec:3} models several runs of the refund department. The different modules of the refund department are introduced in Section~\ref{sec:4}. Proper termination of the system is shown in Section~\ref{sec:5}. Section~\ref{sec:6} verifies the constraints, which are informally introduced in Section~\ref{sec:2}. The paper closes with a discussion of related work and an outlook on further research questions.

\section{The running example: a refund department}\label{sec:2}

The running example of this contribution is a model of the refund procedure of a retail company. The refund department processes customer refund requests (claims). Each claim involves a customer and an item, and is handled by a company staff member, who either approves or rejects it. This process is governed by several constraints:

\begin{enumerate}

\item \emph{Specialized staff}: For each article $a$, there is a designated set of staff, $f(a)$, authorized to handle claims related to $a$. Every incoming request $a$ is assigned to a staff member from the set $f(a)$.

\item \emph{Conflict of interest}: Staff members may also be customers of the retail company. It must be strictly ensured that no staff member processes their own claim.

\item \emph{Limited replacement}: The assigned staff member may be replaced by any other authorized staff member. For each claim, this may happen at most once.

\item \emph{Proper termination}: Each customer claim is ultimately either accepted or rejected.

\end{enumerate}

\begin{figure}[b]
   \centering
   \includegraphics[width=1\textwidth]{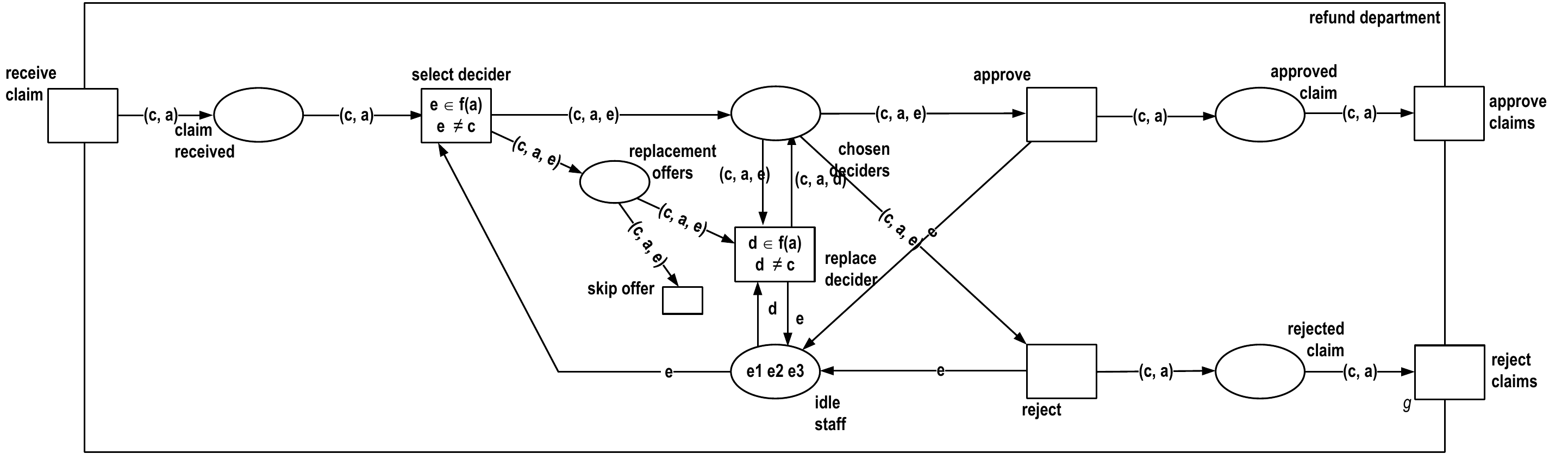}
   \caption{The refund department's bshavior in detail}
   \label{fig:1}
\end{figure}

\begin{figure}[t]
   \centering
   \includegraphics[scale=.4]{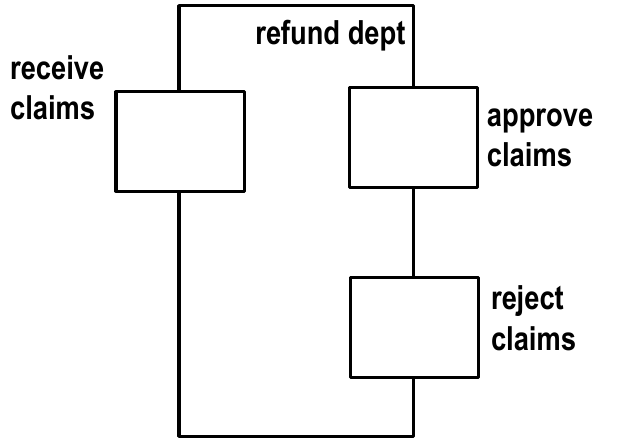}
   \caption{Abstract view on the refund department}
   \label{fig:2}
\end{figure}

\begin{figure}[t]
   \centering
   \includegraphics[width=1\textwidth]{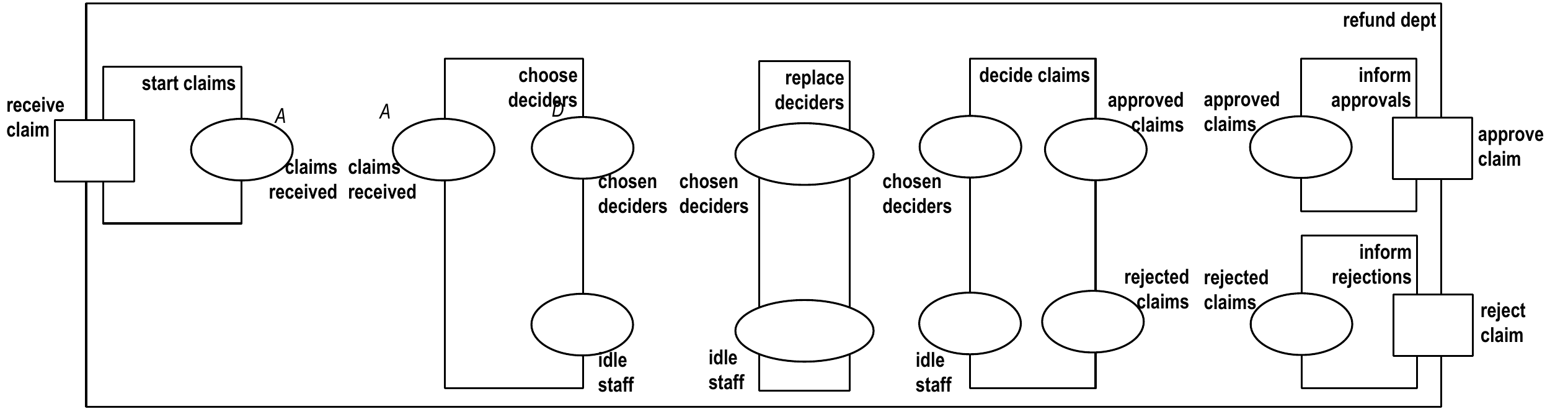}
   \caption{The six modules of the refund department}
   \label{fig:3}
\end{figure}

Fig.~\ref{fig:1} shows a model of the refund department’s behavior. This model is not very understandable; the observer of this model must himself find a structure and an order of the involved events. Furthermore, it is not obvious that the above constraints are satisfied. 

In the sequel, we develop this model clearly, systematically, and in a structured manner that is easy to understand. We start with the most abstract perspective, as shown in Fig.~\ref{fig:2}: Over its left interface, the refund department receives claims; over its right interface, the refund department approves or rejects claims. The boxes indicate that the department interacts with its environment through activities (Petri net transitions), i.e., in handshake mode.

Fig.~\ref{fig:3} refines this model by illustrating the internal structure of the refund department, which includes six modules. The interfaces of these modules include the activities of Fig.~\ref{fig:2}, as well as predicates that may apply to claims, deciders, staff, etc., represented by ellipses. Interestingly, the left and right interfaces of the \emph{replace deciders} module include the same predicates.

\section{Runs (behaviors) of components of the refund department}\label{sec:3}

\begin{figure}[]
   \centering
   \includegraphics[scale=.6]{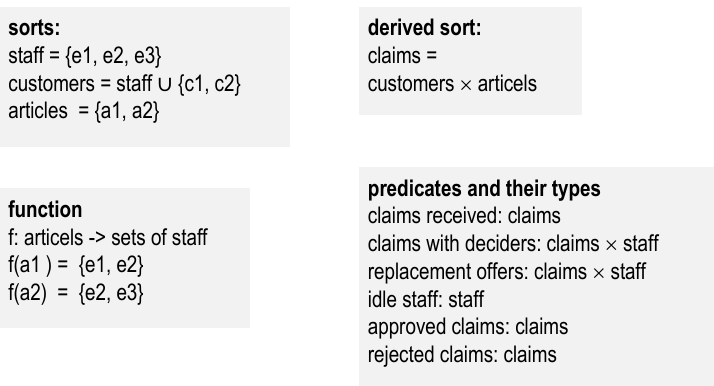}
   \caption{Sets and functions of a given refung procedure}
   \label{fig:4}
\end{figure}

\begin{figure}[]
   \centering
   \includegraphics[width=1\textwidth]{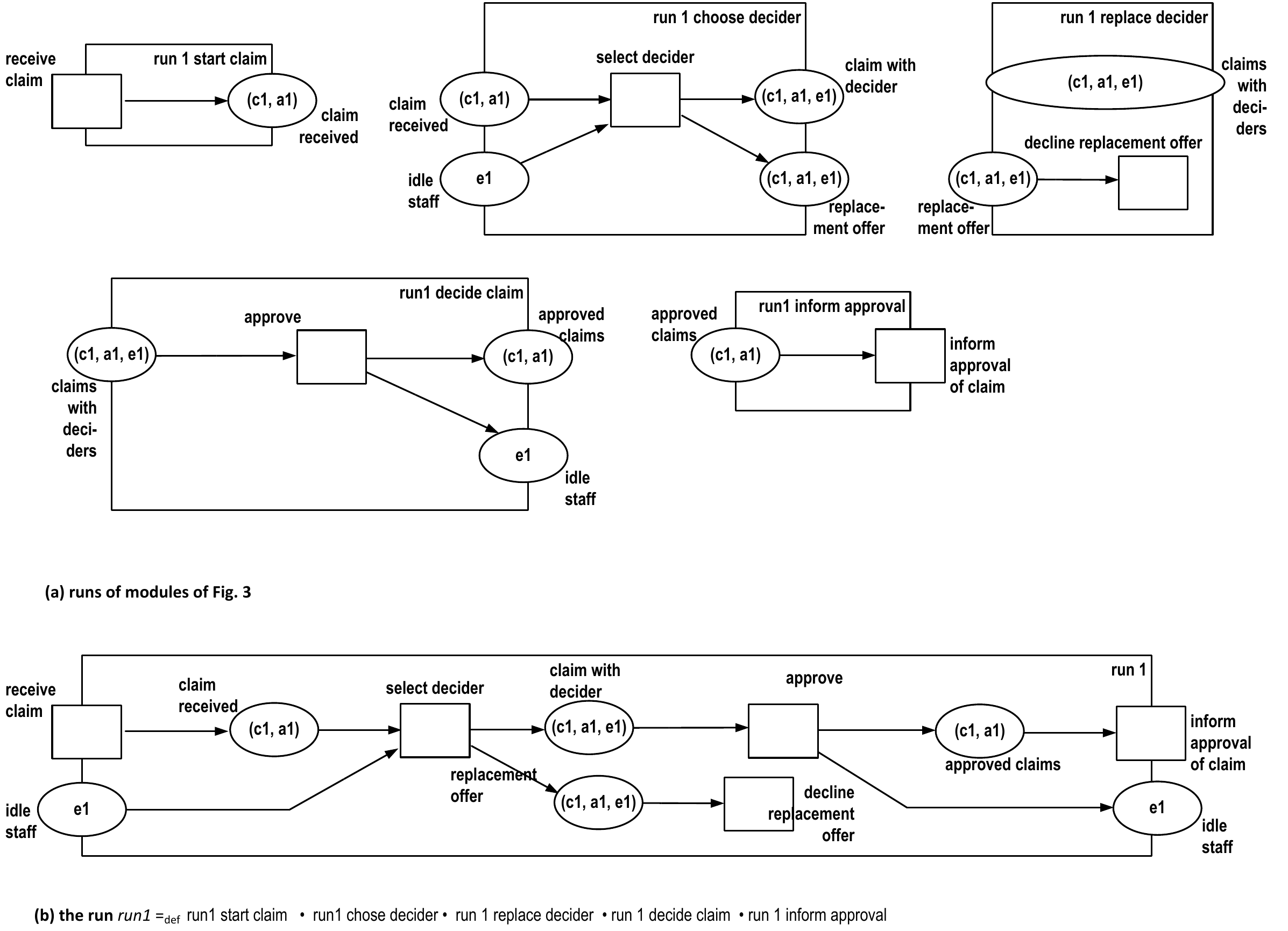}
   \caption{A run of a refund department}
   \label{fig:5}
\end{figure}

Before discussing behavioral aspects, we examine a typical example of a refund department, as shown in Fig.~\ref{fig:4}. There, we have three staff members, five customers, two articles, and a function f that assigns each article a set of staff. 
Classical models of discrete behavior represent single runs (behaviors) as sequences of global states and transitions between these states. The total order of steps then reflects their occurrence over time. However, there are many good reasons to order event occurrences based on \emph{causal} relationships: $b$ is ordered after $a$, in case $b$ can only occur after a has occurred first. As a result, causally independent event occurrences remain unordered. It turns out that this concept aligns very well with both composition and verification. In fact, this is what Petri nets are about.

Fig.~\ref{fig:5}a shows runs of some of the modules of Fig.~\ref{fig:3}: the run $run_1$ of \emph{start claim} receives a claim from customer $c_1$ about article $a_1$. $Run_1$ of \emph{choose decider} selects $e_1$ as a decider for this claim, and offers to replace $e_1$ by some other decider. $Run_1$ of \emph{replace decider} declines this offer and remains with $e_1$. $Run_1$ of \emph{decide claim} approves the claim, and finally, $run_1$ of \emph{inform approval} informs the customer $c_1$.
Fig.~\ref{fig:5}b composes those modules; yielding the run $run_1$:

\begin{equation}
\begin{split}
run_1 \defeq & \textit{ $run_1$ start claim} \compose \textit{$run_1$ chose decider} \compose \\ & \textit{$run_1$ replace decider} \compose \textit{$run_1$ decide claim}  \compose \textit{$run_1$ inform approval}            
\end{split}
\end{equation}

Notice that in $run_1$, the activities \emph{decline replacement offer} occurs independently from \emph{approve} and \emph{inform approval of claim}.

\begin{figure}[t]
   \centering
   \includegraphics[width=1\textwidth]{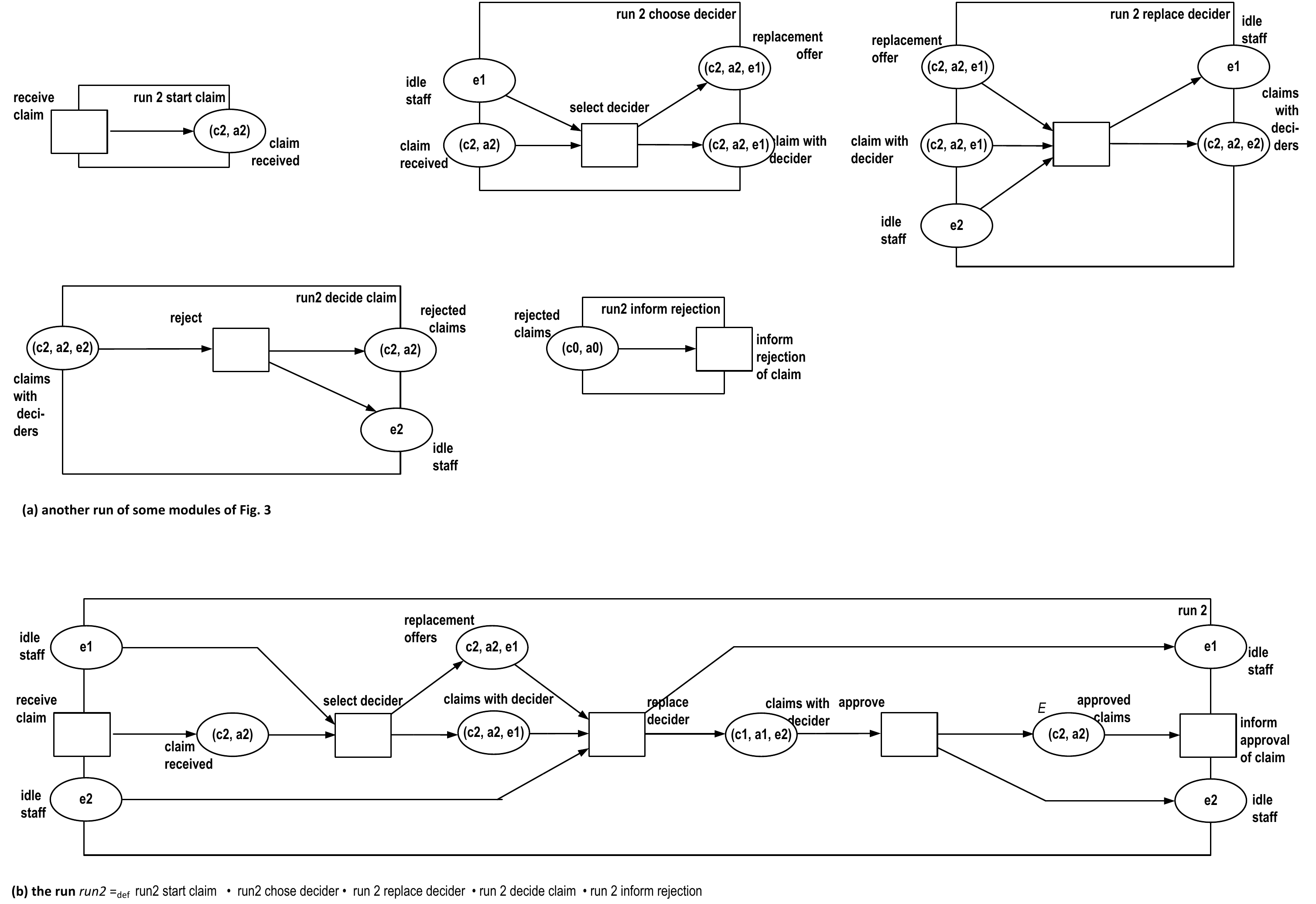}
   \caption{Two other runs and their composition}
   \label{fig:6}
\end{figure}

As a variant of Fig.~\ref{fig:5}, Fig.~\ref{fig:6}a shows runs of the modules of Fig.~\ref{fig:3}, replacing the claim $(c_1, a_1)$ by $(c_2, a_2)$. Only the module \emph{replace decider} operates differently: The decider $e_1$ is replaced with the decider $e_2$. Fig.~\ref{fig:6}b composes these runs, in analogy to Fig.~\ref{fig:5}b:

\begin{equation}
\begin{split}
 run_2 \defeq & \textit{ $run_2$ start claim}   \compose  \textit{$run_2$ chose decider} \compose  \textit{$run_2$ replace decider} \compose \\ & \textit{ $run_2$ decide claim}  \compose  \textit{$run_2$ inform rejection}
\end{split}
\end{equation}

\begin{figure}[t]
   \centering
   \includegraphics[width=1\textwidth]{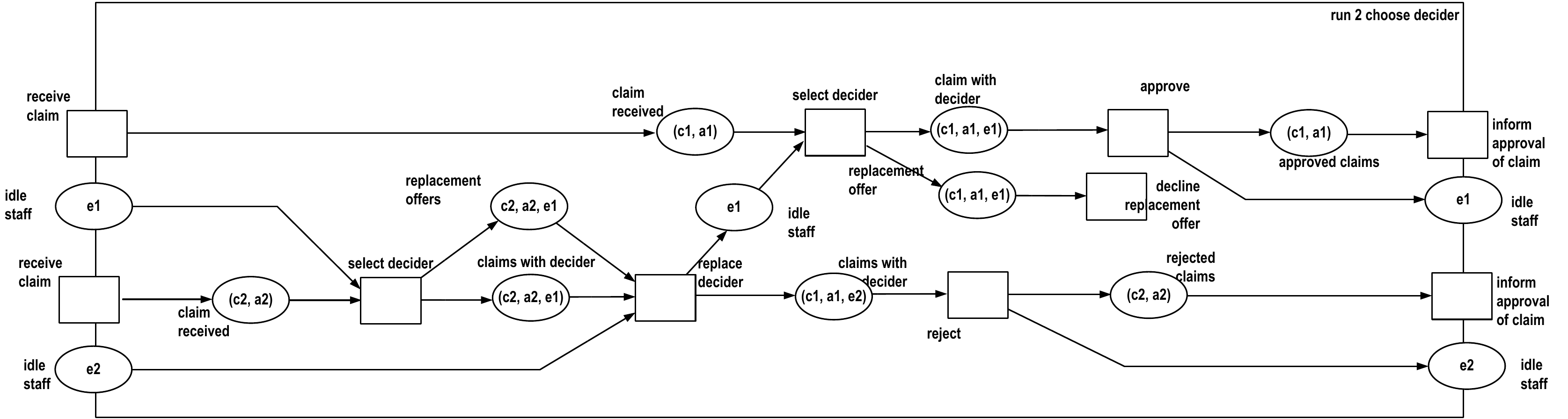}
   \caption{The composition of two runs: $run_2 \compose run_1$}
   \label{fig:7}
\end{figure}

Finally, we can compose the two runs of Fig.~\ref{fig:5}b and Fig.~\ref{fig:5}a, resulting in the run of Fig.~\ref{fig:7}:

\begin{equation}
\begin{split}
run_3 \defeq \textit{ } & run_2 \compose run_1 = \\ &
\textit{$run_2$ start claim}   \compose  \textit{$run_2$ chose decider} \compose  \textit{$run_2$ replace decider} \compose \\ & \textit{$run_2$ decide claim}  \compose  \textit{$run_2$ inform rejection} \compose \\ &
\textit{$run_1$ start claim} \compose \textit{$run_1$ chose decider} \compose \\ & \textit{$run_1$ replace decider} \compose \textit{$run_1$ decide claim}  \compose \textit{$run_1$ inform approval} 
\end{split}
\end{equation}

This run is a run of the system of Fig.~\ref{fig:1}. The order of composition of the modules of the above run is not fixed. For example, \emph{$run_1$ start claim} and \emph{$run_2$ start claim} may be swapped.

\section{A \Heraklit model of components of the refund department}\label{sec:4}

\begin{figure}[t]
   \centering
   \includegraphics[width=1\textwidth]{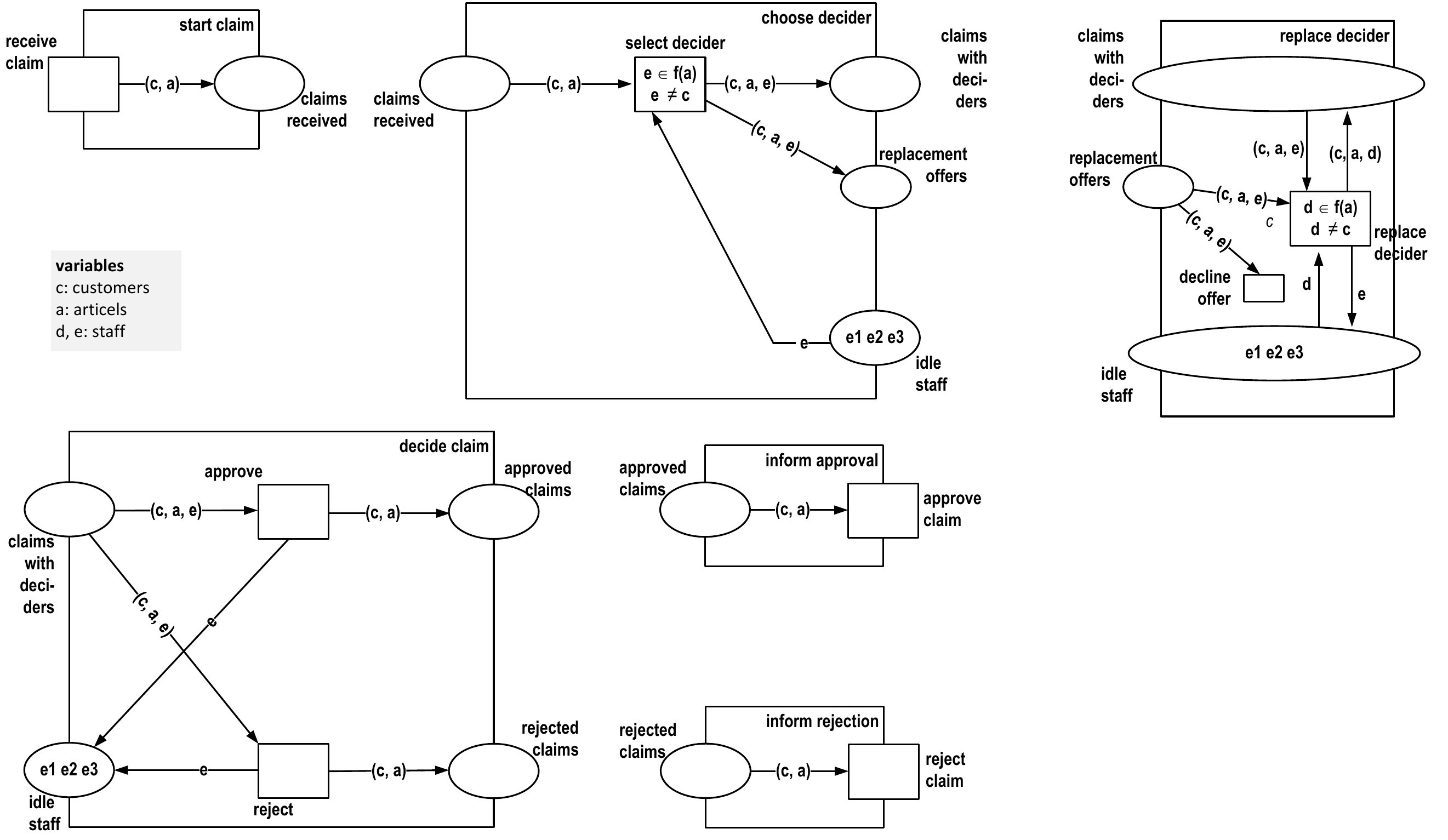}
   \caption{The behavior of the five modules of the refund department}
   \label{fig:8}
\end{figure}

The instance of a refund department as in Fig.~\ref{fig:1} yields already a bunch of runs; Figs.~\ref{fig:5}-\ref{fig:7} show only a few of them. We strive at a finite representation of all of them, in an automaton like fashion. Fig.~\ref{fig:8} shows such a representation as a high-level Petri net. The essential aspect are the variables $c$, $a$, $d$, and $e$. Their valuation with a customer, an article, and staff then yields runs of the modules of Fig.~\ref{fig:4}. For instance, the runs of Fig.~\ref{fig:5} are gained as follows:

\begin{enumerate}

\item Module \emph{start claim}: The transition \emph{receive claim} can occur at any time, instantiating the variables $c$ and a by a concrete client, for instance $c_1$, and a concrete article, for instance $a_1$, yielding the tuple $(c_1, a_1)$ at the predicate \emph{claims received}.

\item Module \emph{choose decider}: With the tuple $(c_1, a_1)$ at the predicate \emph{claims received}, and the authorized staff member $e_1 \in f(a_1)$, occurrence of the transition \emph{select decider} yields the token $(c_1, a_1, e_1)$ at the predicate \emph{chosen decider}, and another copy of this token at the predicate \emph{replacement offers}. Alternatively, the decider $e_2$ or $e_3$ may have been selected.

\item Module \emph{replace decider}: This module substitutes the \emph{chosen decider} $e_1$ with another authorized decider, e.g. $e_2$. Alternatively, the transition \emph{skip offer} eventually revokes the replacement offer.  

\item Module \emph{decide claim}: With the token $(c_1, a_1, e_1)$ at the place chosen decider, both transitions \emph{approve} and \emph{reject} are enabled, but only one of them will occur. Criteria for this choice are not modeled here. 

\item Modules \emph{inform approval} and \emph{inform rejection}: In line with the initial module that receives the claim, these modules pass the decision on the claim to the department’s environment. 

\end{enumerate}

Accordingly, the valuation of the variables $c$ with $c_2$ and a with $a_2$ yields the runs of Fig.~\ref{fig:6}a. In this case, however, the decider $e_1$ is replaced by the decider $e_2$, and the claim is approved.

Composition of the modules of Fig.~\ref{fig:8} then yields the module of Fig.~\ref{fig:1}:

\begin{equation}
\begin{split}
\textit{ refund department} \defeq 
& \textit{ start claim} \compose \textit{ choose decider} \compose \textit{ replace decider} \compose \\ & \textit{ decide claim} \compose \textit{ inform approval} \compose \textit{ inform rejection}
\end{split}
\end{equation}

Summing up, the runs of Fig.~\ref{fig:5}a and \ref{fig:6}a are runs of the system modules in Fig.~\ref{fig:8}. Their respective composition show the Figs.~\ref{fig:5}b, ~\ref{fig:6}b, and \ref{fig:7}.

\begin{figure}[t]
   \centering
   \includegraphics[scale=.6]{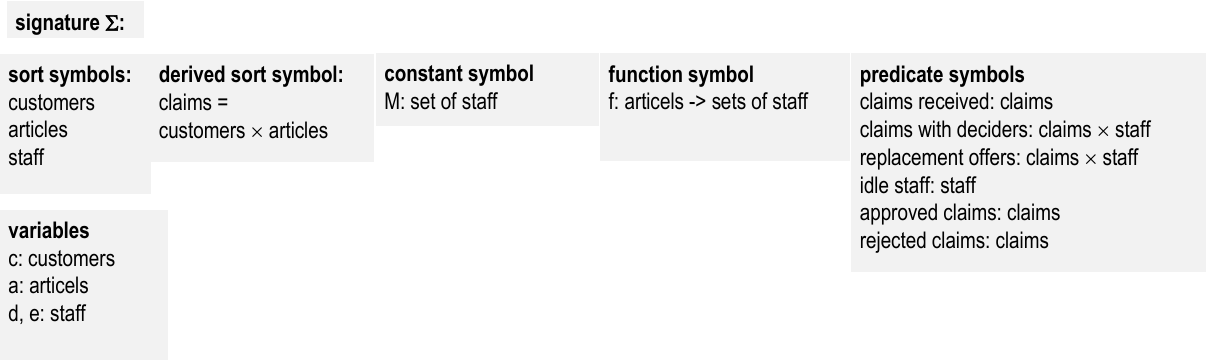}
   \caption{The signature of the refund department}
   \label{fig:9}
\end{figure}

\begin{figure}[t]
   \centering
   \includegraphics[width=1\textwidth]{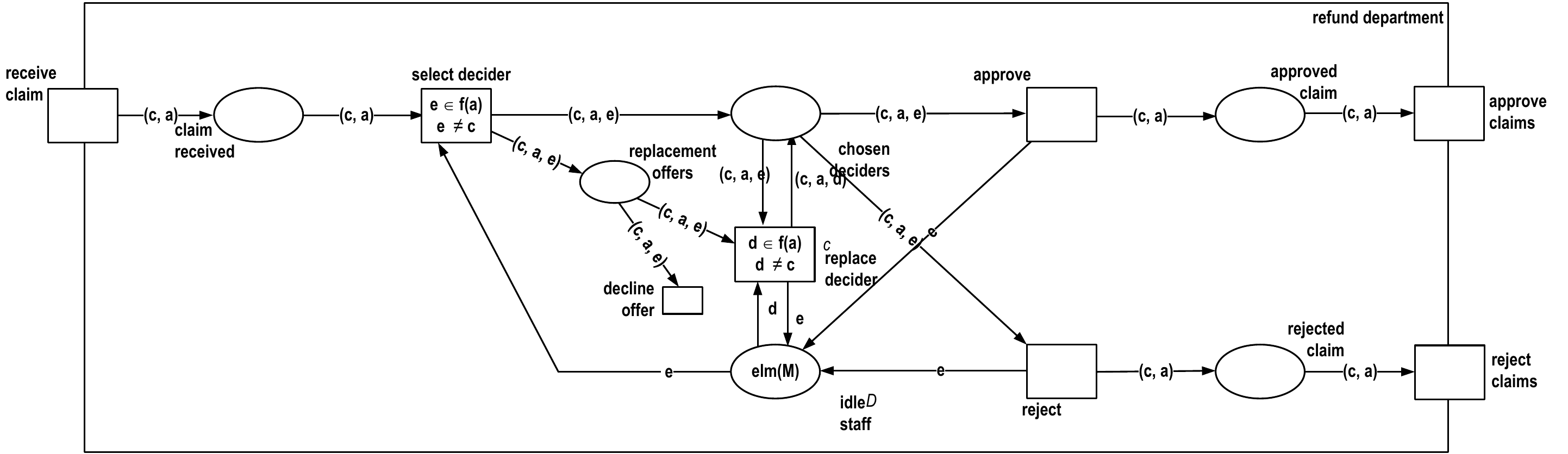}
   \caption{The schema of the refund system}
   \label{fig:10}
\end{figure}

So far, we considered the case of fixed sets of staff, customers and articles, and a fixed function $f$, that assigns each article a set of staff. We may generalize or abstract from these concrete sets and this concrete function, constructing a schema that allows for \emph{any} such set and \emph{any} such function. Technically, we construct a \emph{signature},  in the framework of algebraic structures (in analogy to algebraic specification languages, such as \emph{VDM}, \emph{Z}, etc.). So, the signature $\Sigma$ in Fig. 9 includes symbols for three \emph{sorts}, i.e. symbols for sets of \emph{customers}, \emph{articles}, and \emph{staff}, and a symbol for \emph{claims}. Furthermore, a function symbol $f$ is needed for the function that assigns to each article the eligible staff. The system representation of Fig. 1 almost represents also the schematic version. But there is a decisive problem: The tokens $e_1$, $e_2$, and $e_3$ of the place \emph{idle staff} must be replaced by \say{any staff members}.  A symbol, say, $M$, for a set of staff does not help: considered as a Petri net token, $M$ would be \emph{one} item. Instead, \emph{idle staff} (just as each other place) is a predicate that applies to the elements of (the interpretation of) $M$. We denote this aspect in Fig.~\ref{fig:10} by the inscription \say{$elm(M)$} in the \emph{idle staff} predicate. In formal terms, \say{$elm$} denotes the \say{for all} quantor. This completes the modeling of a refund department for any unspecified company. More on the formal background of the employed modeling technique can be found in \cite{Fettke_Reisig_24}.

\section{Proper termination}\label{sec:5}

\begin{figure}[t]
   \centering
   \includegraphics[width=1\textwidth]{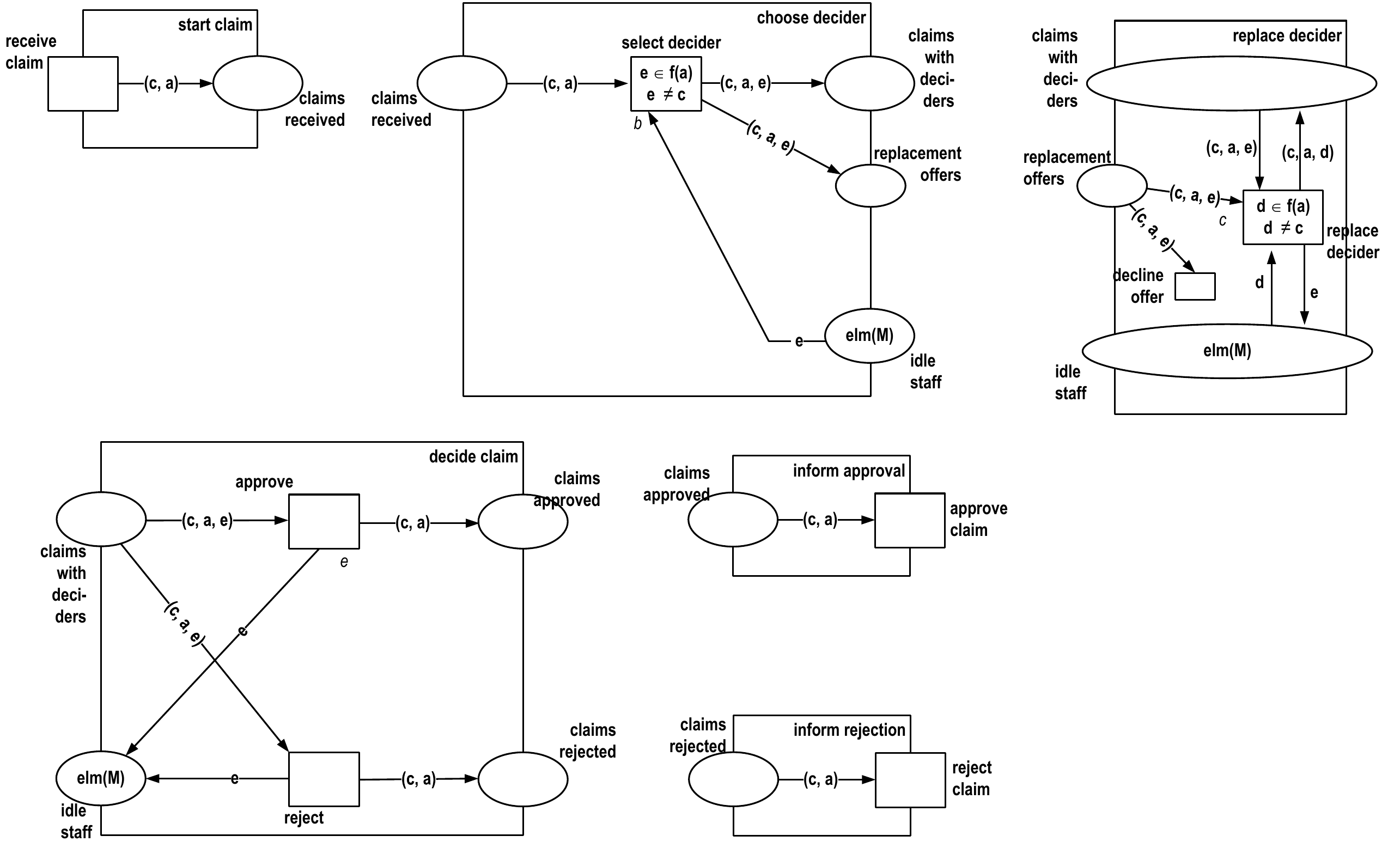}
   \caption{The behavior of the five modules of the refund department}
   \label{fig:11}
\end{figure}

As explained in Sec.~\ref{sec:2}, proper termination requires that each claim is ultimately either accepted or rejected. In technical terms, this means that whenever the transition \emph{receive claim} occurs in a mode $a = a_0$ and $e = e_0$ (with variables $a$ and $e$, an article $a_0$ and a customer $c_0$), eventually one of the transitions \emph{approve claim} or \emph{reject claim} occurs in the same mode. In the framework of the safety/ liveness dichotomy \cite{alpern1985defining}, this is a typical liveness property. Such properties are usually formulated and proven in the framework of temporal logic. 

We suggest a variant of temporal logic, for two reasons. As outlined in Sec.~\ref{sec:3}, we represent single runs as partially ordered sets of events rather than as sequences of events. Second, we suggest composing liveness properties of a composed system from liveness properties of its component modules. Formulated differently, we pick up \say{small} liveness properties directly from the structure of \say{small} modules and compose \say{large} liveness properties from corresponding properties of small modules. We exemplify this in the running example of a refund department, as in Fig.~\ref{fig:11}.   

In technical terms, we employ an operator \say{causes}, written \say{$\mapsto$}, and formulas shaped \say{$p \mapsto q$}, with propositional formulas $p$ and $q$. Propositional formulae and distributed runs are closely related: Each place $p$ is a predicate that applies to the inscribed item a; hence, $p(a)$ is a proposition.  Each transition $t$ is assigned the predicate $occurs(t)$, usually extended by a list of parameters. These parameters correspond to the predicates' parameters in the environment of $t$. Technical details on this can be found in \cite{reisig1999elements}. 

Here we exemplify causes properties along the running example. First, we construct for each of the six modules of Fig.~\ref{fig:8} a cause property. Then we compose these properties, resulting in the wanted formula.

\begin{enumerate}
\item[] The \textbf{start claim} module

\item occurs (receive claim $(c, a)$)     \emph{occurrence rule of Petri nets}

\item occurs receive claim$(c, a)$) $\mapsto$ claims received$(c, a)$   \emph{occurrence rule of Petri nets}

\item \textbf{Start claim} $\models$ occurs (receive claim$(c, a)$) $\mapsto$ claims received$(c, a)$ \emph{definition of the module}

\end{enumerate}

\begin{enumerate}[resume]

\item[] The \textbf{choose decider} module

\item Idle staff.e  $\land$ claims received$(c, a)$ $\mapsto$ claims with deciders$(c, a, e)$ $\land$ replacement offers $(c, a, e)$ \emph{occurrence rule of Petri nets}

\item \textbf{choose decider} $\models$ Idle staff.e  $\land$ claims received(c, a) $\mapsto$ claims with deciders(c, a, e) $\land$ replacement offers $(c, a, e)$ \emph{definition of the module}

\end{enumerate}

\begin{enumerate}[resume]

\item[] The \textbf{replace decider} module

\item idle staff.d  $\land$ claims with deciders$(c, a, e)$ $\land$ replacement offers $(c, a, e)$ $\mapsto$  (claims with deciders$(c, a, e)$ $\land$ idle staff.d) $\lor$ (claims with deciders$(c, a, d)$ $\land$ idle staff.d \emph{occurrence rule of Petri nets}

\item \textbf{replace decider} $\models$ idle staff.d $\land$ claims with deciders$(c, a, e)$ $\land$ replacement offers $(c, a, e)$  $\mapsto$  claims with deciders$(c, a, e)$ \emph{definition of the module}

\end{enumerate}

\begin{enumerate}[resume]

\item[] The \textbf{decide claim} module

\item \textbf{decide claim} $\models$ claims with deciders$(c, a, e)$  $\mapsto$  (claims approved.$(c, a)$) $\lor$ (claimes rejected.$(c, a)$) \emph{occurrence rule of Petri nets}

\end{enumerate}

\begin{enumerate}[resume]

\item[] The \textbf{inform approval} module

\item claims approved.$(c, a)$ $\mapsto$ occurs (approve claim$(c, a)$)  \emph{occurrence rule of Petri nets}

\item \textbf{inform approval} $\models$ claims approved.$(c, a)$ $\mapsto$ occurs (approve claim$(c, a)$)  \emph{definition of the module}

\end{enumerate}

\begin{enumerate}[resume]

\item[] The \textbf{inform rejection} module

\item claims rejected.$(c, a)$ $\mapsto$ occurs (reject claim$(c, a)$)  \emph{occurrence rule of Petri nets}

\item \textbf{inform rejection} $\models$ claims rejected.$(c, a)$ $\mapsto$ occurs (reject claim$(c, a)$)  \emph{definition of the module} 

\end{enumerate}

Proper termination of the return department, as defined in Sec.~\ref{fig:4}, reads

\begin{equation}
\begin{split}
& \textit{occurs} (\textit{receive claim} (c, a)) \mapsto \\ & (\textit{occurs} (\textit{approve claim}(c, a)) \lor \textit{occurs} (\textit{reject claim}(c, a))).
\end{split}
\end{equation}

Validity of this formula is now gained as the composition of the above properties 3., 5., 7., 8., 10., and 12. of the six modules.

\section{Verification of the constraints of Section~\ref{sec:2}}\label{sec:6}

Sec.~\ref{fig:2} presented four requirements for the refund procedure. In fact, the model in Fig.~\ref{fig:11} meets them:

\begin{itemize}

\item \emph{Specialized staff}: This property holds obviously: for each article $a$, the designated set of staff, $f(a)$, is formulated in the structure of Fig.~\ref{fig:4}. At the schematic level, in the signature of Fig.~\ref{fig:9}, is authorized to handle claims related to $a$. Every incoming request $a$ is assigned to a staff member from the set $f(a)$.

\item \emph{Conflict of interest}: There are two transitions that assign a staff member $e$ to a claim: \emph{select decider} and \emph{replace decider}. The inscriptions of both transitions require that customers are never identical with the assigned decider.

\item \emph{Limited replacement}: The assigned staff member may be replaced by any other authorized staff member. For each claim, this may happen at most once. Transition \emph{replace decider} is the only transition that generates replacement. It is geared by the token at place idle.

\item \emph{Proper termination}: This requirement has been proven in Sec.~\ref{sec:5}.

\end{itemize}

\section{Discussion, related work, and conclusions}\label{sec:7}

Computer-integrated systems exhibit two faces: the technological and the applied face. Edsger W. Dijkstra has frequently suggested to strictly separate both sides and to build a \say{firewall} between them \cite{dijkstra:firewallcacm}. His justification: The methods to attack the computer scientists’ formal, mathematical \say{correctness problem} differ fundamentally from the methods to attack the applicants’ informal \say{pleasantness problem}. In this setting, a model is always confined to one side or the other of this wall.

A plethora of work tackles the challenges of solving the correctness and pleasantness problem from different research communities. For example, \say{reactive systems} by Aceto et al. \cite{aceto2007reactive} from a software engineering perspective or \say{Conformance checking} by Carmona et al. \cite{carmona2018conformance} from a BPM perspective. In contrast to Dijkstra and many other approaches, we understand modeling as an activity that should allow a seamless transition between formally and informally given or asserted facts of a computer-integrated system. Technology and applications must be interlocked by shared models, based on the same foundations. As a key concept, our approach is based on a universal idea of composition.

Our case study demonstrates that both sides of Dijskrta's wall can be grounded on the same foundations. It is obvious that this approach has decisive advantages and will achieve tremendous gains: It is seamlessly possible to capture the main ideas of a natural and intuitive understanding of the world we live in, enrich this understanding with formal concepts, and use the description as a foundation for supporting development. As a central feature of our approach, we first verify the properties of the system's components. Based on this verification, the properties of the composite system are formally derived. Future work will expand these ideas into a complete theory.

%
%
%

\bibliographystyle{splncs04}
\bibliography{main}

\end{document}